\documentclass[prl,reprint,amssymb]{revtex4-2}

\usepackage{xcolor}

\usepackage{bm}

\usepackage{amsmath}  
\usepackage{amsfonts}
\usepackage{graphicx}
\usepackage{hyperref}
\hypersetup{
	colorlinks=true,
	citecolor=blue,
	linkcolor=red,
	urlcolor = blue
}

\begin{document}
	
	\title{Charge and spin current pumping by ultrafast demagnetization dynamics}
	
	\author{Jalil Varela-Manjarres}
        \author{Ali Kefayati}
        \author{M. Benjamin Jungfleisch}
        \author{John Q. Xiao}
	\author{Branislav K. Nikoli\'c}
	\email{bnikolic@udel.edu}
	\affiliation{Department of Physics and Astronomy, University of Delaware, Newark, DE 19716, USA}
	
		
\begin{abstract}
The surprising discovery of ultrafast demagnetization---where electric field of femtosecond laser pulse interacts with electrons of a ferromagnetic (FM) layer to cause its magnetization vector {\em to shrink while not rotating}, $M_z(t)/M_z(t=0)<1$---is also assumed to be accompanied by generation of spin current in the direction orthogonal to electric field. However, understanding of the microscopic origin of such spin current, its frequency spectrum and how efficiently it can be converted  into charge current as the putative source of THz radiation, is lacking despite nearly three decades of intense studies. Conversely, quantum transport theory rigorously explains  how microwave driven precession of magnetization vector of {\em fixed} length $\mathbf{M}(t)$ leads to pumping of spin current   into adjacent normal metal (NM) layers sandwiching FM layer to form two-terminal geometry without any applied bias voltage. Here we connect these two apparently disparate phenomena by replacing periodic time-dependence of magnetization precession with nonperiodic time-dependence of demagnetization, as obtained from experiments on ultrafast-light-driven Ni layer, within the same two-terminal setup of standard spin pumping theory. Applying time-dependent nonequilibrium Green's functions, able to evolve such setup with arbitrary time dependence, predicts new phenomenon of charge pumping by demagnetization dynamics, as well as spin, with such currents flowing in directions both parallel and orthogonal to electric field of laser pulse. The pumping of charge current {\em directly} by ultrafast demagnetization dynamics occurs even in the absence of spin-orbit coupling (SOC) and, presumed to be necessary, spin-to-charge conversion via SOC.  Although pumped currents follow $dM_z/dt$ in some setups, this becomes obscured when NM layers are disconnected and pumped currents start to reflect from FM boundaries (as in realistic experimental setups). Finally, we use the Jefimenko equations to compute electromagnetic radiation by charge current pumped in disconnected setup during demagnetization, or later during its slow recovery, unraveling that radiated electric field  {\em only} in the former time interval exhibits features in 
\mbox{0.1--30 THz} frequency range probed experimentally or explored for applications of spintronic THz emitters.
\end{abstract}
	
\maketitle

\begin{figure}
		\centering
		\includegraphics[scale=1.4]{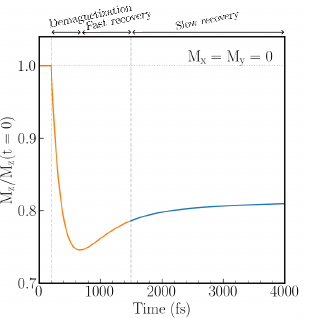}
		\caption{Example of demagnetization dynamics [Eq.~\eqref{eq:demagnetization}] in ultrafast-light-driven single layer of Ni, as extracted by {\em probe} light of time- and angle-resolved photoelectron spectroscopy correlated with time-resolved transverse magneto-optical Kerr effect in Ref.~\cite{Tengdin2018} (reproduced from Fig.~1A in it). The {\em pump} light in Ref.~\cite{Tengdin2018} exciting demagnetization dynamics is fsLP of duration \mbox{$\simeq 28$ fs} and central wavelength  \mbox{$780$ nm}. Note that  nanoscale thickness (\mbox{$\sim 400$ nm}) of Ni film was much greater than both the pumping (\mbox{$\sim 13$ nm}) and probing depth (\mbox{$\sim 1$ nm} for ARPES and \mbox{$\sim 10$ nm} for TMOKE).}
		\label{fig:fig0}
\end{figure}

{\em Introduction}.---The femtosecond (fs)  laser pulse (fsLP)-driven magnetic layer~\cite{Beaurepaire1996} is a far from equilibrium~\cite{Suresh2023,Gillmeister2020} quantum many-body system with very different properties~\cite{Kimel2019} when compared to the same material in equilibrium. It exhibits complex  angular momentum exchange~\cite{Krieger2015,Chen2019a,Chen2019c,Dewhurst2021,Tauchert2022} between photons, electrons, ionic cores and phonons, rapidly emerging  over \mbox{$\sim 10$ fs}  time segments~\cite{Tengdin2018,Siegrist2019}. They conspire to produce ultrafast {\em demagnetization} as experimentally observable phenomenon~\cite{Beaurepaire1996,Tengdin2018,Siegrist2019}, where magnetization vector is decreasing its length along the easy (chosen as the $z$-) axis while {\em not} rotating 
\begin{equation}\label{eq:demagnetization}
	\frac{M_z(t)}{M_z(t=0)}<1; M_x(t)=M_y(t)=0.
\end{equation}
That is, its  $x$- and $y$-components  remain zero or negligible~\cite{Krieger2015,Kefayati2023}, as illustrated by experimental data~\cite{Tengdin2018} in Fig.~\ref{fig:fig0} for single Ni ferromagnetic (FM) layer. Here $M_z(t=0)$ is the magnitude of magnetization in equilibrium, i.e., prior to fsLP  application. Even a single FM layer exhibiting ultrafast demagnetization emits THz electromagnetic (EM) radiation~\cite{Beaurepaire2004}, but such radiation in \mbox{0.1--30 THz} frequency range relevant for applications~\cite{Leitenstorfer2023} becomes greatly enhanced~\cite{Seifert2016,Wu2017,Rouzegar2022} when FM layer is brought into a contact with nonmagnetic (NM) layer hosting strong spin-orbit coupling (SOC)  in the bulk or at the interface~\cite{Jungfleisch2018a,Gueckstock2021,Wang2023}. Experiments also observe~\cite{Kuiper2014} much  faster demagnetization rate in FM/NM  bilayers. 

Although many insights into microscopic mechanisms  causing  demagnetization have been acquired, primarily through first-principles studies based on time-dependent density functional theory (TDDFT)~\cite{Krieger2015,Chen2019a,Chen2019c,Dewhurst2021,Wu2024,Mrudul2024} and its extensions~\cite{Acharya2020,TancogneDejean2018,Barros2022}, the full picture remains incomplete due to rapid accumulation of new experimental data~\cite{Scheid2022,Weissenhofer2024,Tauchert2022} that have to be integrated (such as the role of  phonons~\cite{Tauchert2022,Wu2024}) into TDDFT calculations. Furthermore, even though detecting emission of THz radiation is one of the key experimental  probes~\cite{Gorchon2022} of coupled spin-charge transport in systems exhibiting ultrafast demagnetization, its rigorous  theoretical understanding is lacking. For example, such radiation from a single FM layer has been explained as being of magnetic dipole type, as  emitted by time-dependent total magnetization~\cite{Beaurepaire2004}, but recent detailed TDDFT+Maxwell calculations~\cite{Kefayati2023} show that such 
source is many orders of magnitude weaker than  
time-dependent charges (as it could be expected due to magnetic effects being, in general, $1/c$ smaller than electric ones). In the case of FM/NM bilayers, experiments standardly postulate~\cite{Seifert2016,Wu2017,Rouzegar2022,Seifert2023} the presence of spin current flowing from FM to NM layer, which is then converted into charge current within NM layer by the inverse spin Hall effect (ISHE)~\cite{Saitoh2006}, or at FM/NM interface by other SOC driven mechanisms~\cite{Jungfleisch2018a,Gueckstock2021,Wang2023}. Such charge current within NM layer flows parallel to FM/NM interface, as well as parallel to the initial motion of electrons within FM layer, which follows the direction of the electric field of (typically linearly polarized) of fsLP. 

However, this widely used~\cite{Seifert2016,Wu2017,Rouzegar2022,Seifert2023} picture to interpret experiments, which is plausible~\cite{Gorchon2022} rather than rigorous microscopic one, does not explain: {\em Why FM or FM/NM systems emit EM radiation of frequency so much smaller than that of incoming light? What is the role of demagnetization in it?} The incoming light initially drives~\cite{Krieger2015} valence electrons to respond at its own frequency (typically, fsLP has a central  wavelength of \mbox{$\simeq 800$ nm}), which is expected to lead only to high harmonics (at integer multiples of frequency of incoming light) of emitted radiation as frequently observed  in the case of nonmagnetic materials~\cite{Ghimire2018} (high harmonics are also present~\cite{Suresh2023} in the case of light-driven magnetic materials, but rarely explored experimentally). The usually invoked phenomenological picture of spin voltage (or accumulation)~\cite{Buehlmann2020,Rouzegar2022}, as a difference between nonequilibrium chemical potentials of the two spin species, which drives spin current from FM to NM layer does not explain its frequency spectrum containing features in the THz range or the role played by magnetization varying according to Eq.~\eqref{eq:demagnetization}. For example, the very recent TDDFT study~\cite{Kefayati2023} shows that when electrons respond to light pulse, while magnetization is artificially frozen in time, no spin current flows from FM to NM layer even though spin voltage remains  nonzero in such situation. Thus, time-dependent magnetization [Eq.~\eqref{eq:demagnetization}] as the necessary ingredient to obtain spin current from FM to NM layer {\em points} at it being an additional mechanism, on the top of fsLP, driving quantum subsystem of electrons out of equilibrium.

	\begin{figure}
		\centering
		\includegraphics[scale=1.2]{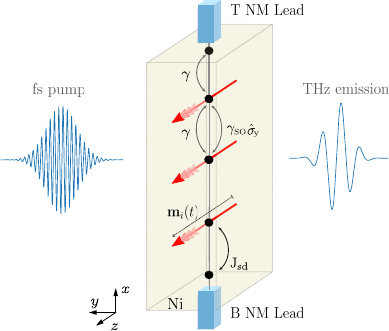}
		\caption{Schematic view of a two-terminal setup---FM central region [modeled by 1D TB chain in Eq.~\eqref{eq:hamil}] connected to semi-infinite $B$ and $T$ NM leads---which is {\em standardly employed} in theoretical studies of spin~\cite{Tserkovnyak2002,Tserkovnyak2005} and charge~\cite{Mahfouzi2012,VarelaManjarres2023} pumping by microwave-driven  magnetization precessing with periodic time dependence. Instead of  precessing magnetization, we use   LMMs  which only change length while not rotating  [Eq.~\eqref{eq:demagnetization}], where such nonperiodic time-dependence of demagnetization is taken from experimental data in Fig.~\ref{fig:fig0}. The setup can also be viewed as 1D chain of atoms isolated from realistic~\cite{Tengdin2018} ultrafast-light-driven Ni layer. Time-dependent $m_i^z(t)$ leads to pumping  of {\em both} charge and spin currents [Fig.~\ref{fig:fig2}], which we compute via TDNEGF algorithms~\cite{Gaury2014,Popescu2016,Petrovic2018,Petrovic2021} that can handle {\em arbitrary} time-dependence within the central region.}
		\label{fig:fig1}
	\end{figure}

 We recall that  mechanisms of charge and spin  pumping~\cite{Citro2023} in time-dependent quantum systems, and in the absence of any DC  bias voltage (hence term ``pumping''), have been amply explored in nanostructures driven by time-dependent gate voltages~\cite{Switkes1999,Brouwer1998}, as well as in  magnetic heterostructures~\cite{Tserkovnyak2002,Tserkovnyak2005,Ando2014a} driven by microwave (or sub-THz in the case of antiferromagentic layers~\cite{Vaidya2020,Dolui2022}) radiation to cause their magnetization into steady precession. In all of these cases, the driving field has periodic time dependence and its frequency is sufficiently small to perturb the system only {\em slightly} out of equilibrium. For example, the energy of microwave photons \mbox{$\hbar \omega \sim 10^{-6}$ eV} is much smaller than the Fermi energy, $\hbar \omega \ll E_F$, so that FM layer with precessing magnetization, acting as periodic time-dependent field that pumps spin current, is in the linear-response regime~\cite{Chen2009}. These problems   offer a blueprint of accomplished fully microscopic understanding, where one starts from time-dependent Hamiltonian (including possible first-principles ones~\cite{Dolui2020b,Dolui2022}) and obtains pumped spin and/or charge currents from  rigorous quantum transport theory~\cite{Brouwer1998,Mahfouzi2012,Moskalets2011,VarelaManjarres2023}. 
 
 On the other hand, the ultrafast-light-driven FM systems appear at first sight quite different from standard systems exhibiting spin pumping~\cite{Tserkovnyak2002,Tserkovnyak2005,Ando2014a} by precessing magnetization of fixed length. The former systems are {\em far from equilibrium} (i.e., with dramatically  modified electronic spectrum~\cite{Suresh2023,Gillmeister2020}) due to $\hbar \omega \sim E_F$, as well as having {\em nonprecessing} [Eq.~\eqref{eq:demagnetization}], {\em nonperiodic} [Fig.~\ref{fig:fig0}] and {\em shrinking in length}  magnetization. Nevertheless, in this Letter we directly connect~\footnote{For prior, but not fully microscopic, attempts to unify low and high energy phenomena in magnets driven out of equilibrium by  EM radiation see Ref.~\cite{Tveten2015}.} these two disparate phenomena while predicting a {\em new effect of  charge pumping by demagnetization dynamics}. We also provide a ``smoking gun'' experiment to confirm it. For this purpose, we adopt an effective field theory strategy~\cite{Burgess2021} by including degrees of freedom required to describe charge current generation and THz radiation by it, while ignoring substructure and degrees of freedom on much shorter length scales. That is, we ignore details~\cite{Krieger2015,Chen2019a,Chen2019c,Dewhurst2021,Wu2024}  of microscopic mechanisms of demagnetization by simply using $\mathbf{M}(t)$ given by experimental data [Fig.~\ref{fig:fig0}].  We employ the same setup from standard spin pumping theory~\cite{Tserkovnyak2002,Tserkovnyak2005,Mahfouzi2012,Dolui2020b,Dolui2022,VarelaManjarres2023} where   FM central region is sandwiched between two semi-infinite NM leads [Fig.~\ref{fig:fig1}]. But we replace precessing localized magnetic moments (LMMs) $\mathbf{m}_i(t)$ at sites $i$ of FM central region (where, e.g., $m_i^z$ is constant and $m_i^x(t)$ and $m_i^y(t)$ change harmonically in time, while $|\mathbf{m}_i(t)|=\mathrm{const.}$~\cite{Tserkovnyak2005,VarelaManjarres2023}) with $m_i^z(t)$ decreasing according to  demagnetization dynamics [Eq.~\eqref{eq:demagnetization}] of experimental data~\cite{Tengdin2018} in Fig.~\ref{fig:fig0}. 

\begin{figure*}
	\centering
	\includegraphics[scale=1.2]{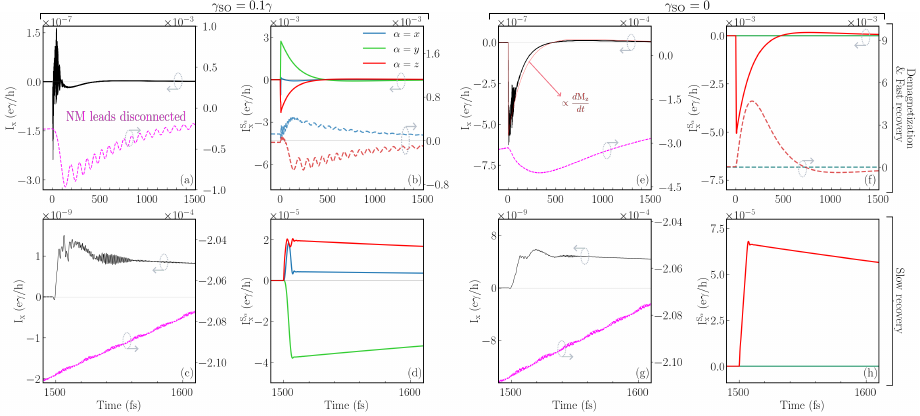}
	\caption{Charge $I_x$ and spin $I_x^{S_\alpha}$ currents along the $x$-axis pumped by demagnetization dynamics [Fig.~\ref{fig:fig0}] of fsLP-driven single Ni 
    layer in: (a)--(d) the presence of SOC coupling [$\gamma_\mathrm{SO} \neq 0$ in Eq.~\eqref{eq:hamil}]; or (e)--(h) the absence of SOC coupling [$\gamma_\mathrm{SO} \neq 0$ in Eq.~\eqref{eq:hamil}]. Panel (e) also shows (thin red line) time derivative $dM_z/dt$. Panels (a),(b),(e) and (f) are obtained during demagnetization \& fast recovery time interval from experimental data in Fig.~\ref{fig:fig0}, while panels (c),(d),(g) and (h) are obtained during slow recovery.  $I_x$ and $I_x^{S_\alpha}$ are computed either in the $T$ NM lead in Fig.~\ref{fig:fig1}; or across the top edge bond ($19 \rightarrow 20$) of FM central region when NM leads are disconnected (dashed lines; for all solid lines, NM leads in Fig.~\ref{fig:fig1} are connected).}
    \label{fig:fig2}
\end{figure*}

Note that the sum of LMMs gives total magnetization, \mbox{$\mathbf{M}=\sum_i \mathbf{m}_i$}, where in the case of standard spin pumping computed via the scattering matrix-based Brouwer formula~\cite{Brouwer1998} one obtains~\cite{Tserkovnyak2002,Tserkovnyak2005} for  the vector of pumped spin current  \mbox{($I^{S_x}_{\mathrm{FM}\rightarrow\mathrm{NM}},I^{S_y}_{\mathrm{FM}\rightarrow\mathrm{NM}},
I^{S_z}_{\mathrm{FM}\rightarrow\mathrm{NM}}) \propto \mathbf{M} \times d\mathbf{M}/dt$}. Na\"{i}ve application of this expression to the setup in Fig.~\ref{fig:fig1} would give zero pumped current as  \mbox{$\mathbf{M} \parallel d\mathbf{M}/dt$} [Eq.~\eqref{eq:demagnetization}] in the case of demagnetization. Nevertheless, since $\mathbf{M}(t)$ is not periodic in the course of demagnetization, neither scattering matrix-based Brouwer formula~\cite{Brouwer1998} nor more general Floquet-scattering-matrix formulas~\cite{Moskalets2011,VarelaManjarres2023} are applicable to two-terminal setup in Fig.~\ref{fig:fig1}. Instead, we employ time-dependent nonequilibrium Green's function (TDNEGF) algorithms~\cite{Gaury2014,Popescu2016} which can handle arbitrary time dependence of the central region in the two-terminal setup of Fig.~\ref{fig:fig1}.  For simplicity, the central FM region in Fig.~\ref{fig:fig1} is modeled as a one-dimensional (1D) tight-binding (TB) chain, and NM leads are modeled as semi-infinite TB chains which terminate at infinity into macroscopic reservoirs of electrons kept at the same chemical potential. This setup can also be viewed [Fig.~\ref{fig:fig1}] as a chain of atoms we isolate from a realistic Ni layer, where incoming linearly polarized laser light with electric field oscillating along the chain (the $x$-axis) causes demagnetization dynamics along the $z$-axis. Such geometry is often encountered experimentally (see, e.g., Fig.~1 in Ref.~\cite{Seifert2023}) or in TDDFT calculations (see, e.g., Fig. 1 in Ref.~\cite{Kefayati2023}). Since experimental setups used in ultrafast demagnetization typically {\em do not} include NM leads, reservoirs and external circuit attached---in contrast to standard  pumping problems  studied in nanoscale devices~\cite{Switkes1999,Citro2023} or magnetic multilayers~\cite{Ando2014a}---we also analyze an additional setup in Fig.~\ref{fig:fig1} whose semi-infinite NM leads are disconnected (due to very small hopping toward them) and pumped currents (dashed lines in Fig.~\ref{fig:fig2})  are forced to reflect from the boundaries of the FM central region. 

\begin{figure}[b!]
	\centering
	\includegraphics[scale=0.22]{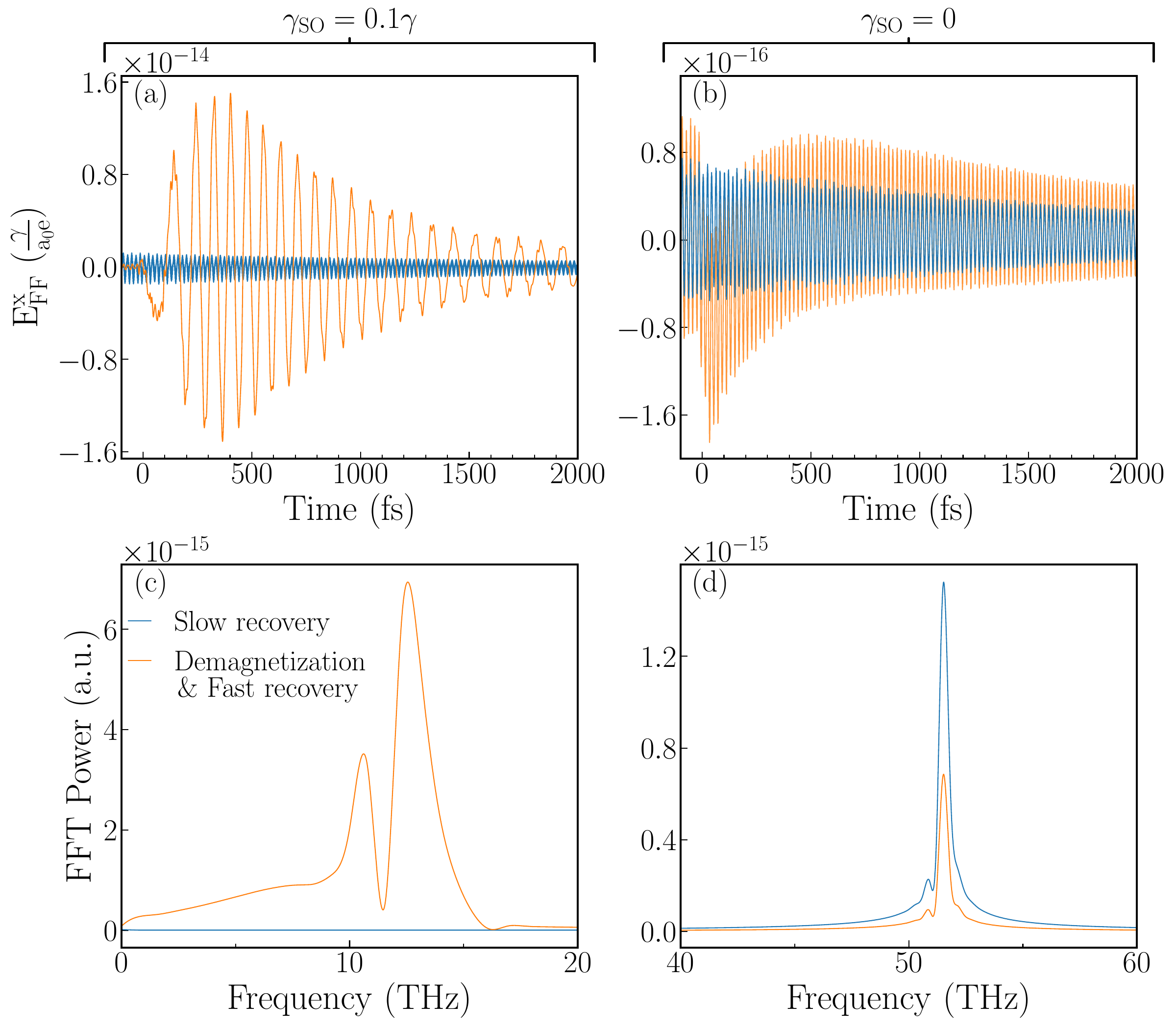}
	\caption{(a) Time-dependence of the $x$-component of electric field $E^x_\mathrm{FF}$ of EM radiation in the far field region, as generated by pumped charge bond currents from Figs.~\ref{fig:fig2}(a) [orange line] and ~\ref{fig:fig2}(c) [blue line] in the presence of SOC. (b) Same information as in  (a), but using pumped charge bond currents from Figs.~\ref{fig:fig2}(e) and Figs.~\ref{fig:fig2}(g) in the absence of SOC. (c) and (d)  FFT power of signals in panels (a) and (b), respectively.}
\label{fig:fig3}
\end{figure}

Our principal results in Figs.~\ref{fig:fig2} and ~\ref{fig:fig3} divulge what kind of currents can be pumped by demagnetization dynamics, as well as the frequency spectrum of EM radiation due to time-derivative of pumped charge current [Eq.~\eqref{eq:efield}], respectively. Prior to delving into the results, we introduce useful notation and basic concepts of time-dependent quantum transport and EM radiation computation via the Jefimenko formulas~\cite{Jefimenko1966,McDonald1997}.

{\em Models and Methods}.---The Hamiltonian of two-terminal setup of Fig.~\ref{fig:fig1} is that of 1D TB chain 
	\begin{eqnarray}\label{eq:hamil}
		\hat{H}(t) & = & - \gamma \sum_{\langle ij \rangle}  \hat{c}_{i}^{\dagger}  \hat{c}_{j}  - J_{sd}\sum_i\hat{c}_i^\dagger \hat{\bm \sigma} \cdot \mathbf{m}_i(t) \hat{c}_i \nonumber \\ 
		\mbox{} && - i\gamma_\mathrm{SO} \sum_{\langle ij \rangle}  \hat{c}_{i}^{\dagger}  \hat{\sigma}_y \hat{c}_{j}.
	\end{eqnarray}
    The TB chain hosts both conduction electrons and classical LMMs $\mathbf{m}_i(t)$ to model metallic  FM central region. The bottom ($B$) and the top ($T$) NM leads are also semi-infinite 1D TB chains, described by the first term alone in Eq.~\eqref{eq:hamil}. The Fermi energy of the macroscopic reservoirs into which NM leads terminate is set at $E_F=0$. Here   \mbox{$\hat{c}_i^\dagger=(\hat{c}_{i\uparrow}^\dagger \  \ \hat{c}_{i\downarrow}^\dagger)$} is a row vector containing operators $\hat{c}_{i\sigma}^\dagger$ which create an electron with spin $\sigma=\uparrow,\downarrow$ at site $i$; $\hat{c}_i$ is a column vector containing the corresponding annihilation operators; $\gamma$ is the hopping between the nearest-neighbor (NN) sites, also setting the unit of energy; $\gamma_\mathrm{SO}$ is an additional spin-dependent hopping~\cite{Nikolic2006} due to the Rashba SOC~\cite{Manchon2015}; and the conduction electron spin, described by the vector of the Pauli matrices \mbox{$\hat{\bm  \sigma} = (\hat{\sigma}_x,\hat{\sigma}_y,\hat{\sigma}_z)$}, interacts with $\mathbf{m}_i(t)$, which remain parallel at all times to the $z$-axis while only changing their length [Eq.~\eqref{eq:demagnetization}] according to experimental data in Fig.~\ref{fig:fig0},  via $sd$ exchange interaction of strength \mbox{$J_{sd} = 0.2\gamma$}~\cite{Cooper1967}.  
    
	The fundamental quantity of quantum statistical mechanics is the density matrix. The time-dependent one-particle nonequilibrium density matrix  can be expressed~\cite{Gaury2014},  ${\bm \rho^{\rm neq}}(t) = \hbar \mathbf{G}^<(t,t)/i$, in terms of the lesser Green's function of TDNEGF formalism~\cite{Gaury2014}  defined by \mbox{$G^{<,\sigma\sigma'}_{ii'}(t,t')=\frac{i}{\hbar} \langle \hat{c}^\dagger_{i'\sigma'}(t') \hat{c}_{i\sigma}(t)\rangle_\mathrm{nes}$} where  
	$\langle \ldots \rangle_\mathrm{nes}$ is the nonequilibrium statistical average~\cite{Stefanucci2013}. We solve a matrix integro-differential equation~\cite{Popescu2016,Petrovic2018}  
	\begin{equation}\label{eq:rhoneq}
		i\hbar \partial_t {\bm \rho}^{\rm neq} = [\mathbf{H}(t),{\bm \rho}^{\rm neq}] + i \sum_{ p= B,T} [{\bm \Pi}_p(t) + {\bm \Pi}_p^{\dagger}(t)],
	\end{equation}
	for the time evolution of ${\bm \rho}^{\rm neq}(t)$, where $\mathbf{H}(t)$ is the matrix representation of Hamiltonian in Eq.~\eqref{eq:hamil} and $\partial_t \equiv \partial/\partial t$. Equation~\eqref{eq:rhoneq} is an {\em exact} quantum master equation for the reduced density matrix of the FM central  region viewed as an open finite-size quantum system attached to macroscopic Fermi liquid reservoirs via semi-infinite NM leads.   The ${\bm \Pi}_p(t)$ matrices
	\begin{equation}\label{eq:current}
		{\bm \Pi}_p(t) = \int_{0}^t \!\! dt_2\, [\mathbf{G}^>(t,t_2){\bm \Sigma}_p^<(t_2,t) 
		- \mathbf{G}^<(t,t_2){\bm \Sigma}_p^>(t_2,t) ],
	\end{equation} 
	are expressed in terms of the lesser and greater Green's functions~\cite{Stefanucci2013} and the corresponding self-energies ${\bm \Sigma}_p^{>,<}(t,t')$~\cite{Popescu2016}. They yield  directly time-dependent charge, $I_p(t)  =  \frac{e}{\hbar} \mathrm{Tr}\, [{\bm \Pi}_p(t)]$, and
	and spin current, $I_p^{S_{\alpha}}(t) = \frac{e}{\hbar} \mathrm{Tr}\, [\hat{\sigma}_{\alpha}{\bm \Pi}_p(t)]$, flowing into lead $p = B,T$ for {\em arbitrary} time-dependence of Hamiltonian of the central region. 	Since the  applied bias voltage between NM leads is identically zero in this study, all computed $I_p(t)$ and 
	$I_p^{S_{\alpha}}(t)$ are solely currents pumped by time-dependence of the Hamiltonian. We use the same units for charge and spin currents, defined as \mbox{$I_p = I_p^{\uparrow} + I_p^{\downarrow}$} and \mbox{$I_p^{S_{\alpha}} = I_p^{\uparrow} - I_p^{\downarrow}$}, in terms of spin-resolved charge currents $I_p^{\sigma}$. In our convention, positive current in NM lead $p$ means charge or spin current is flowing out of that lead. 

     The electric field of EM radiation emitted into the FF region is calculated from the Jefimenko equations~\cite{Jefimenko1966},  reorganized~\cite{McDonald1997} to isolate the contribution in the far field (FF) region (where electric field decays as inverse of distance from the source)
\begin{widetext}
\begin{equation}\label{eq:efield}
\mathbf{E}_\mathrm{FF}(\mathbf{r}, t)=\frac{1}{4\pi \epsilon_0 c^2}\sum_{P_{i \rightarrow j}=1}^{N_b} \int_{P_{i \rightarrow j}}\bigg[ (\mathbf{r}-\mathbf{l})\frac{\partial_t I_{i \rightarrow j}(t_r)}{|\mathbf{r}-\mathbf{l}|^3}(\mathbf{r}-\mathbf{l})\cdot \mathbf{e}_x - \frac{\partial_t I_{i \rightarrow j}\left(t_r\right)}{|\mathbf{r}-\mathbf{l}|} \mathbf{e}_x \bigg ] d l,
\end{equation}
\end{widetext}
     as well as  adapted~\cite{Ridley2021,Suresh2023} to take time-dependent bond charge currents, $I_{i \rightarrow j}(t)$ [Eq.~\eqref{eq:bondcharge}]  defined on TB lattice, as the source. Note that Jefimenko equations can also be viewed~\cite{Griffiths1991} as proper time-dependent generalizations of the Coulomb and Biot-Savart laws. Here,  \mbox{$t_r \equiv t -|\mathbf{r}-\mathbf{l}|/c$} emphasizes retardation in the response time due to relativistic causality~\cite{Jefimenko1966,McDonald1997}. In Figs.~\ref{fig:fig2} and ~\ref{fig:fig3}, as well as in  Eq.~\eqref{eq:efield}, we use  $N=20$ as the number of TB sites and $N_b=19$ as the number of bonds between.  The bond currents~\cite{Nikolic2006,Petrovic2018,Ridley2021} $I_{i\rightarrow j }$ are assumed to be spatially
homogeneous along the path $P_{i\rightarrow j}$ from site $i$ to site $j$, which
is composed of a set of points $l \in P_{i\rightarrow j}$. We obtain them as
     \begin{equation}\label{eq:bondcharge}
		I_{i\rightarrow j}(t) = \frac{e\gamma}{i\hbar}\mathrm{Tr}_{\rm spin}\bigg[{\bm \rho}_{ij}(t)\mathbf{H}_{ji}(t) - {\bm \rho}_{ji}(t)\mathbf{H}_{ij}(t)\bigg],
	\end{equation}
     by isolating  $2 \times 2$ submatrices	${\bm \rho}^\mathrm{neq}_{ij}(t)$  of ${\bm \rho}^\mathrm{neq}(t)$ whose off-diagonal elements determine such currents. Note that diagonal elements of ${\bm \rho}^\mathrm{neq}_{ij}(t)$ determine  on-site nonequilibrium charge density, whose time dependence contributes~\cite{Ridley2021,Suresh2023,Kefayati2023} to near-field EM  radiation. In Eq.~\eqref{eq:bondcharge} $\mathrm{Tr}_{\rm spin}[\ldots]$ denotes trace in the spin space only.

{\em Results and Discussion}.---Figure~\ref{fig:fig2}(e)--(g) demonstrates that demagnetization dynamics pumps {\em both} charge and (one component of, $I^{S_z}_x$) spin currents, even in the absence of any SOC within FM central region . This is quite {\em surprising} when compared to standard pumping by precessing magnetization where charge pumping is found  only under special conditions, such as SOC present in the bulk~\cite{VarelaManjarres2023} of FM layer, or at FM/NM interface~\cite{Mahfouzi2012}, as confirmed experimentally~\cite{Ciccarelli2015}. With SOC absent and NM leads attached, we also find [Figs.~\ref{fig:fig2}(e),(f)] that during demagnetization pumped currents are  $\propto dM_z/dt$. This provides some justification for the so-called ``$dM/dt$ mechanism'' conjectured~\cite{Choi2014,Lichtenberg2022,Rouzegar2022} from fitting of experimental data. However, with leads disconnected and/or in the presence of SOC---which is typically the case as SOC  provides magnetic anisotropy in FM layers in equilibrium or  plays an essential role~\cite{Krieger2015,Siegrist2019} out of equilibrium by triggering spin-flips~\cite{Acharya2020} responsible for demagnetization---pumped currents proportionality to $dM_z/dt$ is lost  in Fig.~\ref{fig:fig2}. In addition, all three components of pumped spin current [Figs.~\ref{fig:fig2}(b),(d)] become nonzero in the presence of SOC. 

Time-dependent charge currents [Figs.~\ref{fig:fig2}(a),(e)]  will inevitably radiate EM waves, even if conversion of additionally pumped spin current [Figs.~\ref{fig:fig2}(b),(f)]  into charge current is: absent, as in case of a single FM layer~\cite{Beaurepaire2004}; inefficient~\cite{Gorchon2022}; or not easily related~\cite{Schmidt2023} to the source of experimentally observed THz radiation. We use the Jefimenko equations~\cite{Jefimenko1966,McDonald1997}, as properly time-retarded solutions of the Maxwell equations in the case when time-dependent charges and their current can be considered as given~\cite{Kefayati2023}, to compute electric field $\mathbf{E}_\mathrm{FF}$ [Eq.~\eqref{eq:efield}] of emitted EM radiation in the FF region. The Jefimenko formula for $\mathbf{E}_\mathrm{FF}$ reveals that it is radiated only by the {\em time-derivative} of local [or bond, $I_{i \rightarrow j}$, in Eqs.~\eqref{eq:efield} and ~\eqref{eq:bondcharge}] charge currents, rather than by charge current itself  as often assumed in fitting of experimental data~\cite{Rouzegar2022,Seifert2023}. By plugging into Eq.~\eqref{eq:efield} bond charge currents from realistic setup with disconnected NM leads, i.e.,  magenta curves from Figs.~\ref{fig:fig2}(a) and ~\ref{fig:fig2}(e), we obtain $E_\mathrm{FF}^x$ in Figs.~\ref{fig:fig3}(a) and ~\ref{fig:fig3}(b), respectively. Their fast Fourier transform (FFT) in Figs.~\ref{fig:fig3}(c) and ~\ref{fig:fig3}(d), respectively, demonstrates that charge currents pumped directly (i.e., {\em without} any spin-to-charge conversion) by demagnetization dynamics contain [orange line in Fig.~\ref{fig:fig3}(c)] spectral features within \mbox{0.1--30 THz} range in full accord with experiments~\cite{Seifert2016,Wu2017,Rouzegar2022,Seifert2023,Jungfleisch2018a,Gueckstock2021,Wang2023}. By comparing Figs.~\ref{fig:fig3}(c) and ~\ref{fig:fig3}(d), we also learn that THz spectral features are  related to SOC-induced oscillations in magenta line in Fig.~\ref{fig:fig2}(a).


{\em Conclusions and Outlook}.---By using time dependence [Fig.~\ref{fig:fig0}] of ultrafast demagnetization [Eq.~\eqref{eq:demagnetization}] from experiments~\cite{Tengdin2018} on fsLP-driven Ni layer---which is plugged into a two-terminal setup [Fig.~\ref{fig:fig1}] of standard theory~\cite{Tserkovnyak2002,Tserkovnyak2005} of spin pumping to replace its slowly and harmonically precessing magnetization of fixed length driven by  microwave absorption~\cite{Ando2014a}---we {\em directly connect} these two apparently  disparate phenomena. That is, time-dependent quantum transport theory that can handle~\cite{Gaury2014,Popescu2016,Petrovic2018,Petrovic2021} arbitrary time-dependence of LMMs within the central FM region in Fig.~\ref{fig:fig1} shows how  demagnetization pumps, surprisingly~\cite{Mahfouzi2012,VarelaManjarres2023,Ciccarelli2015}, both spin and charge currents. The physical picture emerging is that fsLP drives electrons far from equilibrium to cause their current oscillating at light frequency (as well as at high harmonics of light frequency~\cite{Ghimire2018,Suresh2023}), while the ensuing~\cite{Krieger2015} demagnetization dynamics pumps {\em additional} charge current whose time derivative [Eq.~\eqref{eq:efield}] generates spectral features of emitted EM radiation at THz frequencies in Fig.~\ref{fig:fig3}(c). Additional Fig.~S1 in the Supplemental Material~\footnote{See Supplemental Material at \url{https://wiki.physics.udel.edu/qttg/Publications}, which includes Ref.~\cite{Bajpai2019}, for two additional figures. Figure~S1 re-plots orange curves from Figs.~\ref{fig:fig3}(a) and ~\ref{fig:fig3}(c), obtained when only demagnetization dynamics drives electrons out of equilibrium, and compares them to their  counterparts when both fsLP and demagnetization dynamics are present or just fsLP is present. Figure~S2 plots the same information as in Fig.~\ref{fig:fig3}(c) while varying parameters in the Hamiltonian [Eq.~\eqref{eq:hamil}] or number of TB sites.} demonstrates that presence of {\em both} fsLP and demagnetization dynamics, as concurrent nonequilibirum drives, does not change our conclusions from  
 Fig.~\ref{fig:fig3}(c).  We also find that slow recovery [Fig.~\ref{fig:fig0}] of magnetization on longer  \mbox{$\sim 1$ ps} timescales  {\em does not} generate any EM radiation with features in \mbox{0.1--30 THz} range [Fig.~\ref{fig:fig3}(d)]. Our prediction of direct charge pumping by ultrafast demagnetization dynamics offers a unified explanation for experimentally observed THz radiation from  both single FM layer~\cite{Beaurepaire2004} and FM/NM bilayers~\cite{Seifert2016,Wu2017,Rouzegar2022,Seifert2023}. It can be easily {\em confirmed or falsified} by observing THz radiation from, e.g., Pt/Ni/Pt trilayer whose intensity  is comparable to the one from often  employed~\cite{Seifert2023} Ni/Pt bilayer. In contrast, standard phenomenological picture~\cite{Seifert2016,Wu2017,Rouzegar2022,Seifert2023} of spin-to-charge current conversion by ISHE in Pt would  {\em predict no} EM radiation from Pt/Ni/Pt trilayer due to opposite direction of ISHE charge currents within two Pt layers. Even if much smaller THz radiation is found in Pt/Ni/Pt trilayer, our theory still provides microscopic explanation for the origin of spin current flowing from Ni to Pt layer, thereby replacing phenomenologically conjectured spin voltage~\cite{Buehlmann2020,Rouzegar2022} as its  driving mechanism. 

\begin{acknowledgments}
This research was primarily supported by NSF through the  University of Delaware Materials Research Science and Engineering Center, DMR-2011824. The supercomputing time was provided by DARWIN (Delaware Advanced Research Workforce and Innovation Network), which is supported by NSF Grant No. MRI-1919839.
\end{acknowledgments}


\bibliography{references}

\end{document}